\documentclass{emulateapj}
\RequirePackage{lineno}
\usepackage{amssymb}
\usepackage{amsmath}
\usepackage{mathrsfs}
\usepackage{ulem}
\usepackage{booktabs}
\usepackage{longtable}
\usepackage{natbib}
\usepackage[colorlinks, linkcolor=blue, urlcolor=blue, citecolor=blue]{hyperref}
\usepackage{array}
\usepackage{float}
\usepackage{graphicx}
\usepackage{subfigure}
\usepackage{color}
\usepackage[all]{hypcap}
\usepackage{multirow}
\usepackage[toc,page]{appendix}
\usepackage{makecell}

\def\beq{\begin{equation}}
\def\enq{\end{equation}}
\def\bea{\begin{eqnarray}}
\def\ena{\end{eqnarray}}

\def\apjl{ApJL\,}

\begin{document}

\title{The Prompt Emission of Gamma-Ray Bursts from the Wind of Newborn Millisecond Magnetars: A Case Study of GRB 160804A}
\author{Di Xiao\altaffilmark{1,2}, Zong-kai Peng\altaffilmark{1,2}, Bin-Bin Zhang\altaffilmark{1,2}, and Zi-Gao Dai\altaffilmark{1,2}}
\affil{\altaffilmark{1}School of Astronomy and Space Science, Nanjing University, Nanjing 210093, China; dxiao@nju.edu.cn; bbzhang@nju.edu.cn; dzg@nju.edu.cn}
\affil{\altaffilmark{2}Key Laboratory of Modern Astronomy and Astrophysics (Nanjing University), Ministry of Education, China}

\begin{abstract}
In this paper, we revisit the scenario that an internal gradual magnetic dissipation takes place within the wind from a newborn millisecond magnetar can be responsible for gamma-ray burst production. We show that a combination of two emission components in this model, i.e., the photospheric emission from the wind and the synchrotron radiation within the magnetic reconnection region, can give a reasonable fit to the observed spectrum of the prompt emission phase of GRB 160804A. We obtain the physical parameters through a Monte Carlo procedure and deduce the initial spin period and magnetic field of the central magnetar. Furthermore, the independent afterglow fitting analysis gives a consistent result, adding great credibility to this scenario. In addition, we predict a subclass of GRBs called bursts from such a Magnetar wind Internal Gradual MAgnetic Dissipation (abbreviated as ``MIGMAD bursts'') that have several distinctive properties.

%We also perform a Monte Carlo procedure to obtain the best fitting parameters and discuss their implications on the properties of the central engine.
\end{abstract}


\section{Introduction}

Millisecond magnetars are widely believed to be one of the possible central engines of gamma-ray bursts \citep[for recent reviews, see][]{fab12,kum15,dai17}, either from the collapse of massive stars or double neutron star mergers. After formation, a proto-magnetar ejects an outflow and produces a gamma-ray burst (GRB). Generally, for a magnetar central engine, there are two kinds of outflow. One is driven due to accretion of the magnetar and the other is the proto-magnetar wind. The former extracts the gravitational energy of accreted matter and can reach a high luminosity to power a normal GRB that is similar to a black hole central engine\citep[e.g.][]{zhang08,zhang09,zhang10} while the latter extracts the stellar rotational energy and is usually less luminous \citep{uso92, dai98a,dai98b}. There are at least two indirect observational evidences for the existence of such a wind. First, a pulsar wind nebula is suggested to originate from the shock produced as the pulsar wind interacts with an interstellar medium. For instance, the measured radio spectrum of the Crab Nebula is naturally explained if a wind with a Lorentz factor of $\sim10^4$ from the Crab pulsar is introduced \citep{ato99}. Second, a bunch of GRB X-ray afterglows exhibit plateau or flare features that can be well explained by the long-term energy injection from central engines \citep[e.g.][]{zhang06}. The injected energy is possibly from the spin-down power that is mediated by the wind\footnote{ Note that there are some instances that are not in favor of magnetar's energy injection \citep{benia17}. The other possible kind of injected energy comes from the fall-back accretion of the black hole system \citep[e.g.][]{ruf97,ros03,lei13,wu13}. However, the fall-back mass may be too large explaining the plateau or flare at very late times ($>10^6\,\rm s$) \citep[for a recent review, see][]{liu17}.} \citep[e.g.][]{yu09,yu10,lv18}. This proto-magnetar wind is initially cold and Poynting-flux dominated \citep{met11}. Later, as the magnetic energy being converted to the kinetic energy of bulk motion, this wind turns to be electron-positron dominated that can further energize the interstellar medium. In this process, very high-energy gamma-rays can be produced, as confirmed by the {\it Fermi}-LAT observation of Crab pulsar \citep{aha12}. Motivated by this, we would like to study the high-energy
emission from the magnetic energy dissipation in the wind of the newborn magnetar.

The mechanism of GRB prompt emission has not been well known. In the standard model, it is attributed to the synchrotron emission from a bunch of accelerated electrons by internal shocks \citep{ree94, kob97, daig98}. Also, acceleration could be realized via abrupt magnetic reconnection \citep{zhang11}. However, the synchrotron model usually suffers from the low energy index problem, i.e., the low energy photon index ($\alpha$) fitted by the Band function is often found close to $-1$ or even larger. Therefore we need additional physics to explain the observed spectrum. Possible approaches include involving marginally slow cooling regime of radiating electrons \citep{kum08, daig11, ben13, ben18}, inverse Compton scattering \citep{bos09, daig11, gen18} or decay of a magnetic field behind the shock \citep{der07, lem13, uhm14}. Alternatively, the dissipative photospheric model can reproduce $\alpha\sim-1$ spectrum if dissipative mechanisms (e.g. internal shocks, magnetic reconnection, etc) occur just below the photosphere radius \citep{tho94, ghi99, ree05, pee05, bel10, gia12, lun13, den14, beg15, gao15}. Another scenario involves the internal gradual magnetic dissipation mechanism which suggests that the magnetic energy of a highly magnetized outflow can be converted to thermal energy and bulk kinetic energy via gradual dissipation \citep{spr01, dre02a, dre02b, gia05, met11, gia12, ben17b}. In the meantime, magnetic reconnection is able to accelerate electrons and produce non-thermal radiation \citep{ben14}. Observationally, one would expect superposition of thermal and non-thermal components that may lead to a hard low-energy spectral index \citep{ben17}. Based on this consideration, in this paper we revisit the scenario that involves magnetic dissipation within the wind of a newborn magnetar as a possible mechanism for producing GRBs, which is named the Magnetar wind Internal Gradual MAgnetic Dissipation model (abbreviated as MIGMAD model and hereafter).

%focus on the third model and try to explain the observational properties.

This paper is organized as follows. In Section 2 we introduce the MIGMAD model and predict its emission. Then, we perform a comprehensive case study of GRB 160804A and discuss some implications of its central magnetar in Section 3. At last, we present our conclusions in Section 4.

\section{The MIGMAD model}

A newborn millisecond magnetar is one of the possible central engines of both long and short GRBs. A wind from this proto-magnetar should be squeezed and collimated into a narrow jet while propagating in the envelope material, and finally penetrate out since the wind is long-lasting comparing to its break-out time \citep{bro11, bro14, bro16}. This wind is initially Poynting-flux dominated \citep{cor90} and its magnetic energy can be converted to thermal emission and bulk kinetic energy via internal gradual magnetic dissipation \citep{spr01, dre02a, dre02b, gia05}. In the meantime, electrons can be accelerated by reconnection. Particle-in-Cell (PIC) simulations suggest that these accelerated electrons could have a power-law distribution with an index $p$ \citep{sir14, guo15, kag15, wer16}. Therefore, synchrotron emission is also expected from these electrons \citep{ben14, ben17}. We calculate each component in this section.

At a given radius, the Poynting-flux luminosity of the wind could be written as \citep{gia05, ben17}
\beq
L_B=c\frac{(rB)^2}{4\pi}=L_0\left[1-\frac{\Gamma(r)}{\Gamma_{\rm sat}}\right],
\enq
where $L_0$ is the total luminosity of the wind (per steradian), and $B$ and $\Gamma(r)$ are the magnetic field strength and Lorentz factor of the wind at radius $r$ respectively. $\Gamma_{\rm sat}$ is the bulk Lorentz factor of the wind at the saturation radius that is given by $r_{\rm sat}=\lambda\Gamma_{\rm sat}^2/(6\epsilon)=1.7\times10^{15}\Gamma_{\rm sat,4}^2(\lambda/\epsilon)_8\,\rm cm$ \citep{ben17}, where the typical ``wavelength" of the field is $\lambda=cP=3\times10^7P_{-3}\,\rm cm$ in the striped wind configuration \citep{cor90, spr01, dre02a, dre02b} and $\epsilon\sim0.1-0.25$ is the ratio of the reconnection velocity to the speed of light \citep{guo15, liu15}. Since the co-moving temperature decreases as $T^{\prime}\propto r^{-7/9}$, the thermal luminosity decreases as $L_{\rm th}(r)\propto r^{-4/9}$ \citep{gia05}, substituting the energy dissipation rate $d\dot{E}=-(dL_B/dr)dr$, then the total thermal photospheric luminosity can be obtained by integrating from the initially launching radius to the photospheric radius $r_{\rm ph}$ \citep{gia05, ben17, xiao17b},
\bea
L_{\rm ph}&=&\int_0^{r_{\rm ph}}\frac{1}{2}\left(\frac{r}{r_{\rm ph}}\right)^{4/9}d\dot{E}\nonumber\\
&=&2.6\times10^{47}L_{0,50}^{6/5}\Gamma_{\rm sat,4}^{-1}\left(\frac{\lambda}{\epsilon}\right)_8^{-1/5}\,\rm erg\,s^{-1}\,sr^{-1},
\ena
with the temperature being
\beq
T_{\rm ph} = 95L_{\rm 0,50}^{1/10}\Gamma_{\rm sat,4}^{1/4}\left(\lambda\over \epsilon\right)_8^{-7/20}\,{\rm keV}\label{Tph},
\enq
where $r_{\rm ph}$ can be obtained by setting the Thomson scattering depth $\tau(r_{\rm ph})=1$, which gives
$r_{\rm ph}=3.0\times10^9L_{0,50}^{3/5}\Gamma_{\rm sat,4}^{-1}(\lambda/\epsilon)_8^{2/5}\,\rm cm$ \citep{ben17,xiao17b}. So the thermal component can be written as
\beq
L_{\nu}^{\rm ph}(\nu)= L_{\nu}^{\rm ph}(\nu;L_0, \lambda/\epsilon, \Gamma_{\rm sat}).
\enq

In order to obtain the synchrotron spectrum, we need to calculate the relevant break frequencies. The three break frequencies $\nu_m,\,\nu_c,\,\nu_a$ depend on radius and can be obtained by the same way as in \citet{xiao17b}. $\nu_{\max}$ is the maximum frequency corresponds to the maximum electron Lorentz factor that is determined by equaling the synchrotron energy loss timescale with reconnection acceleration timescale. Initially the electrons are in the fast cooling regime for which the spectrum is \citep{sar98}
\beq
L_\nu^{\rm syn}=\begin{cases}
L_{\nu,\max}^{\rm syn}(\nu/\nu_c)^{1/3} & \text{if}\,\,\nu<\nu_c, \\
L_{\nu,\max}^{\rm syn}(\nu/\nu_c)^{-1/2} & \text{if}\,\,\nu_c<\nu<\nu_m, \\
L_{\nu,\max}^{\rm syn}(\nu_m/\nu_c)^{-1/2}(\nu/\nu_m)^{-p/2} & \text{if}\,\,\nu_m<\nu<\nu_{\max},
\end{cases}
\enq
where
\begin{equation}
L_{\nu,\max}^{\rm syn}=\frac{m_ec^2\sigma_T\Gamma B^{\prime}N_e(r)}{3q},
\end{equation}
with $N_e(r)$ being the total number of emitting electrons in the wind at $r$.
Letting $\nu_m=\nu_c$, we can get the radius $r_{\rm tr}$ at which the transition from fast cooling to slow cooling happens. Since usually $\nu_a>\nu_c$ holds at $r_{\rm ph}$, we can define another critical radius $r_{\rm cr}$ at which $\nu_a$ crosses $\nu_c$. Then the whole synchrotron spectrum can be written as follows \citep{xiao17b}. Initially for $r_{\rm ph}<r\leq r_{\rm cr}$,
\beq
L_\nu^{\rm syn}=\begin{cases}
L_{\nu_a}^{\rm syn}(\nu/\nu_a)^{11/8} & \text{if}\,\,\nu<\nu_a, \\
L_{\nu_a}^{\rm syn}(\nu/\nu_a)^{-1/2} & \text{if}\,\,\nu_a<\nu<\nu_m, \\
L_{\nu_a}^{\rm syn}(\nu_m/\nu_a)^{-1/2}(\nu/\nu_m)^{-p/2} & \text{if}\,\,\nu_m<\nu<\nu_{\max}.
\end{cases}
\enq
Further for $r_{\rm cr}\leq r\leq r_{\rm tr}$,
\beq
L_\nu^{\rm syn}=\begin{cases}
L_{\nu_a}^{\rm syn}(\nu/\nu_a)^{11/8} & \text{if}\,\,\nu<\nu_a, \\
L_{\nu_a}^{\rm syn}(\nu/\nu_a)^{1/3} & \text{if}\,\,\nu_a<\nu<\nu_c, \\
L_{\nu_a}^{\rm syn}(\nu_c/\nu_a)^{1/3}\\ \,\,\times (\nu/\nu_c)^{-1/2} & \text{if}\,\,\nu_c<\nu<\nu_m, \\
L_{\nu_a}^{\rm syn}(\nu_c/\nu_a)^{1/3}\\ \,\,\times (\nu_m/\nu_c)^{-1/2}(\nu/\nu_m)^{-p/2} & \text{if}\,\,\nu_m<\nu<\nu_{\max}.
\end{cases}
\enq
Lastly, for $r_{\rm tr}\leq r\leq r_{\rm sat}$,
\beq
L_\nu^{\rm syn}=\begin{cases}
L_{\nu_a}^{\rm syn}(\nu/\nu_a)^{11/8} & \text{if}\,\,\nu<\nu_a, \\
L_{\nu_a}^{\rm syn}(\nu/\nu_a)^{1/3} & \text{if}\,\,\nu_a<\nu<\nu_m, \\
L_{\nu_a}^{\rm syn}(\nu_m/\nu_a)^{1/3}\\ \,\,\times (\nu/\nu_m)^{-(p-1)/2} & \text{if}\,\,\nu_m<\nu<\nu_c, \\
L_{\nu_a}^{\rm syn}(\nu_m/\nu_a)^{1/3}\\ \,\,\times (\nu_c/\nu_m)^{-(p-1)/2}(\nu/\nu_c)^{-p/2} & \text{if}\,\,\nu_c<\nu<\nu_{\max}.
\end{cases}
\enq

The non-thermal synchrotron spectrum can be obtained by integrating the above expressions from the photospheric radius to the saturation radius. So the non-thermal component can be given by
\beq
L_{\nu}^{\rm syn}(\nu)=L_{\nu}^{\rm syn}(\nu;L_0, \lambda/\epsilon, \epsilon_e, \Gamma_{\rm sat}, p)
\enq
where $\epsilon_e$ is the fraction of dissipated energy per electron.

Lastly, the total spectrum is the sum of the thermal and non-thermal fluxes,

%\bea
%F_{\nu}(\nu)&=&F_{\nu}(\nu;\beta,\xi,\zeta,\omega,p,\rm Norm) \nonumber\\
%&=&\frac{L_{\nu}^{\rm ph}(\nu;\beta,\xi,\omega)+L_{\nu}^{\rm syn}(\nu;\beta,\xi,\zeta,\omega,p)}{4\pi D_L^2}\times \rm Norm,\nonumber\\
%\ena

\bea
F_{\nu}(\nu)&=&F_{\nu}(\nu;L_0, \lambda/\epsilon, \epsilon_e, \Gamma_{\rm sat}, p) \nonumber\\
&=&\frac{L_{\nu}^{\rm ph}(\nu;L_0, \lambda/\epsilon, \Gamma_{\rm sat})+L_{\nu}^{\rm syn}(\nu;L_0, \lambda/\epsilon, \epsilon_e, \Gamma_{\rm sat}, p)}{4\pi D_L^2}\nonumber\\
\ena
where $D_L$ is the luminosity distance.

\section{Application to GRB 160804A}

GRB 160804A is a long gamma-ray burst that triggered both {\it Fermi}-GBM and {\it Swift}-BAT. It has a duration of $T_{90}\sim130\,\rm s$ and peak energy $E_p \sim 100 \,\rm keV$ \citep{bis16, mar16}, which match the expectations of the MIGMAD model well. The afterglow emission of GRB 160804A is discovered by several groups and the absorption and emission line features in the optical afterglow suggests a redshift of $z\simeq0.736$ \citep{xu16}. In particular, the X-ray afterglow monitored by {\it Swift}-XRT exhibited a prominent shallow decay phase \citep{dav16}, which adds credibility to a magnetar central engine. In addition, the burst duration during which the observed (gamma-ray and X-ray) flux is dominated by jet emission is $t_{\rm burst}\simeq10^3\,\rm s$ \citep{zhang14,gao17}, which is also in favor of a long-lasting magnetar inside. Therefore, we consider GRB 160804A as a perfect candidate and we will discuss in detail here.

\subsection{Prompt Emission of GRB 160804A}

GRB 160804A triggered {\it Fermi}-GBM at $T_0=01:34:16.04$ UT on 04 August 2016 and the public data is processed using a pipeline described in \cite{zhang16}. Since the prompt phase lasts over 100 seconds, we need to do a time-resolved spectral analysis. As a trial, we binned the light curve into ten time intervals equally, starting from $T_0-50\,\rm s$ to $T_0+100\,\rm s$. In each time interval, the spectrum is preliminarily fitted by the Band function and the evolutions of its parameters ($\alpha,\,\beta,\,E_p$) are shown in Figure 1.
We can see that $\alpha$ is close to $-1$ and $E_p$ is several tens of keV in most time intervals. We note that the last four intervals are dominated by background photons so the Band function parameters are unconstrained. We thus only focus on the first six time intervals in our analysis.

%\begin{figure}
%\label{fig1}
%\begin{center}
%\includegraphics[width=0.45\textwidth]{fig1.eps}
%\caption{The prompt photon counts of GRB 160804A in different energy bands. ``n8,\,n4,\,b0" stand for three {\it Fermi}-GBM detectors.}
%\end{center}
%\end{figure}

\begin{figure}
\label{fig1}
\begin{center}
\includegraphics[width=0.5\textwidth]{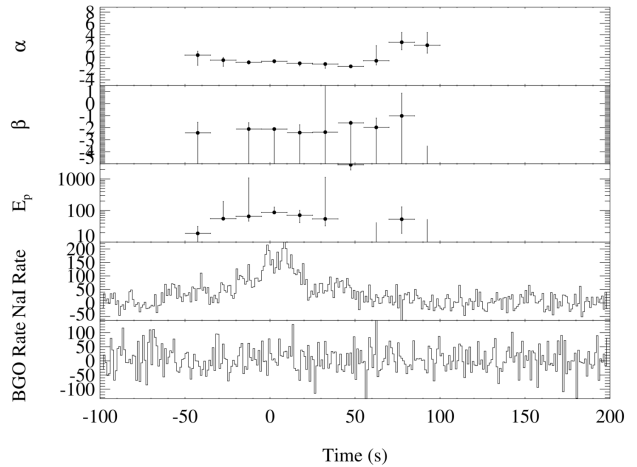}
\caption{The time-resolved spectrum fitting results with the Band function. The top three panels show the evolution of three parameters $\alpha,\,\beta,\,E_p$ defined in the Band function during ten time intervals. The bottom two panels show the light curves of NaI and BGO detectors on board {\it Fermi}. }
\end{center}
\end{figure}

In order to test the MIGMAD model, we need to interpret the prompt emission properties of GRB 160804A.
We fit the first six time intervals using the MIGMAD model with McSpecFit package \citep{zhang16}. Spectral fitting is then performed within the allowed range via a Bayesian Monte-Carlo method. In a typical manner, Figure 2 shows the parameter corner plot of interval-3 (main peak of the light curve). The spectrum fitting results are shown in Figure 3 and 4, while the best-fitting parameters are listed in Table 1. As we can see clearly, the MIGMAD model fits the observed spectrum well, with PGSTAT/d.o.f close to unity. Furthermore, thermal components that are predicted in the MIGMAD model are clearly seen in all six intervals. Moreover, we can plot the time evolutions of these five parameters in Figure 5, which give a lot of information about the central magnetar that will be discussed later in section 3.3.

\begin{figure*}
\label{fig2} \centering\includegraphics[angle=0,height=7.5in]{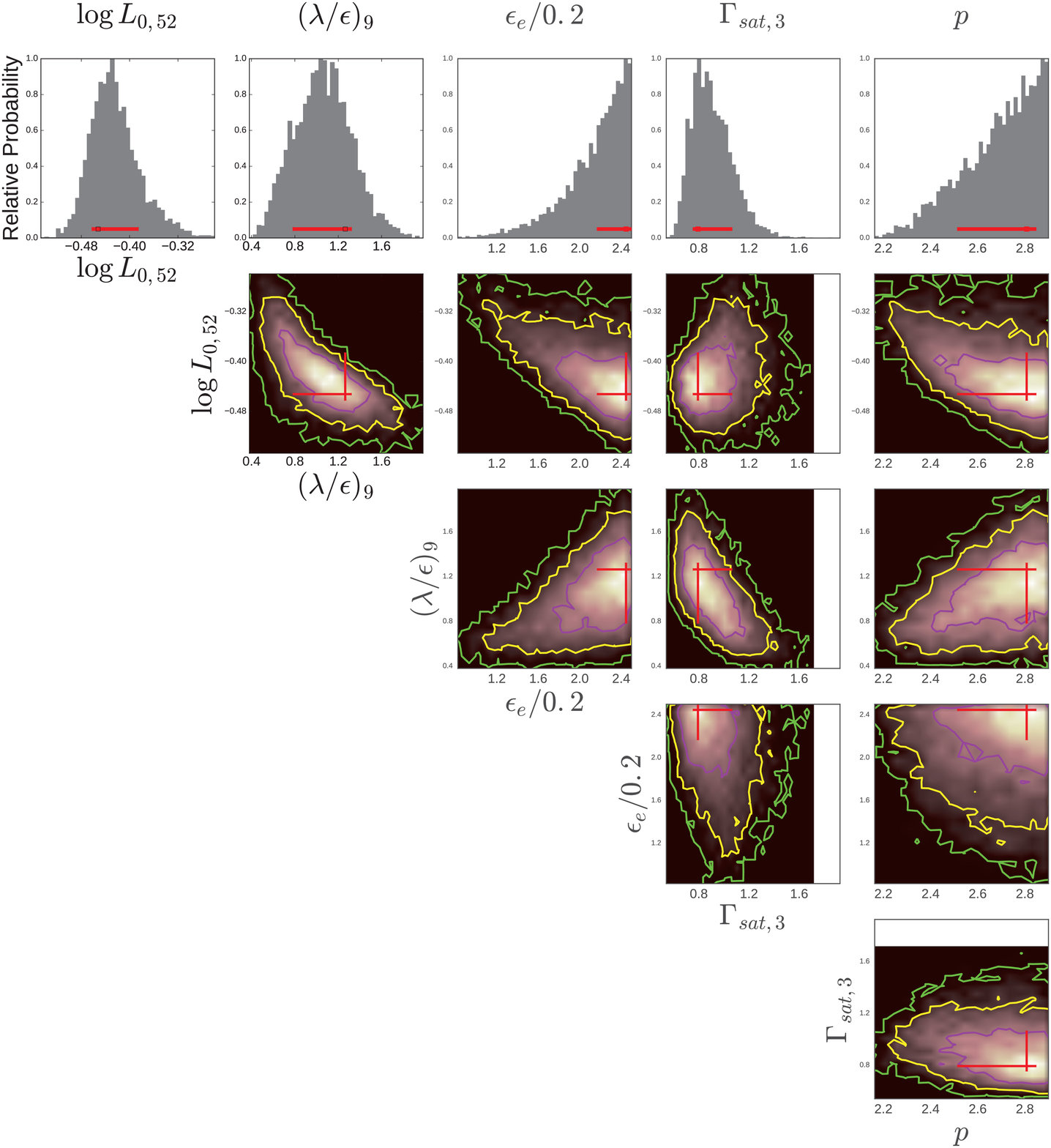} \ \
\caption{Parameter constraints of the MIGMAD model fitting for the spectrum in time interval-3 (from -5 s to 10 s). Histograms and contours illustrate the likelihood map. Red crosses show the best-fitting values and
1-sigma error bars.}
\end{figure*}

%\begin{figure*}
%\label{fig4} \centering\includegraphics[angle=0,height=4.0in]{fig4a.eps} \
%\ \centering\includegraphics[angle=0,height=4.0in]{fig4b.eps} \ \
%\caption{Spectrum fitting results. \textit{Left panel}: \ The observed count spectrum and model count %spectrum for GRB 160804A. \textit{Right panel}:
%The observed photon flux (black datapoints) and theoretical photon spectrum (red line).}
%\end{figure*}

\renewcommand\arraystretch{2}
\begin{table*}
\centering
\caption{Best-fitting parameters with the MIGMAD model for six time-intervals of GRB160804A}
\vspace{2mm}
%\resizebox{\textwidth}{26mm}{
\begin{tabular}{c|c|c|c|c|c|c|c}
\hline
\hline
\multirow{3}{*}{Parameter} & \multirow{3}{*}{Allowed range} & \multicolumn{6}{c}{Best-fitting value} \\ \cline{3-8}
& &
interval-0 & interval-1 & interval-2 & interval-3 & interval-4 & interval-5 \\
& & [-50 s, -35 s] &[-35 s, -20 s] &[-20 s, -5 s] &[-5 s, 10 s] &[10 s, 25 s] &[25 s, 40 s] \\
\hline
$\log L_{0,52}$ & [-5, 0] & $-1.19_{-0.19}^{+0.12}$ & $-0.61_{-0.18}^{+0.08}$ & $-0.82_{-0.01}^{+0.13}$ & $-0.45_{-0.01}^{+0.07}$ & $-0.63_{-0.12}^{+0.13}$ & $-0.81_{-0.10}^{+0.21}$ \\
\hline
$(\lambda/\epsilon)_9$ & [0.1, 3.0] & $2.91_{-0.64}^{+0.09}$ & $2.71_{-0.79}^{+0.29}$ & $2.50_{-0.64}^{+0.50}$ & $1.26_{-0.48}^{+0.06}$ & $1.24_{-0.18}^{+0.95}$ & $2.15_{-0.64}^{+0.54}$\\
\hline
$\epsilon_e/0.2$ &[0.25, 2.5] & $0.94_{-0.09}^{+1.22}$ & $0.70_{-0.20}^{+0.68}$ & $2.44_{-0.42}^{+0.06}$ & $2.44_{-0.28}^{+0.05}$ & $1.46_{-0.28}^{+0.84}$ & $1.18_{-0.27}^{+1.01}$\\
\hline
$\Gamma_{\rm sat,3}$ & [0.1, 2.0] & $0.32_{-0.08}^{+1.22}$ & $1.49_{-0.92}^{+0.12}$ & $0.51_{-0.21}^{+0.26}$ & $0.79_{-0.04}^{+0.27}$ &$1.04_{-0.21}^{+0.36}$ & $0.82_{-0.40}^{+0.40}$ \\
\hline
$p$ & [2.1, 2.9] & $2.32_{-0.08}^{+0.46}$ & $2.11_{-0.01}^{+0.28}$ & $2.83_{-0.21}^{+0.07}$ & $2.81_{-0.29}^{+0.04}$ & $2.61_{-0.36}^{+0.18}$ & $2.87_{-0.26}^{+0.03}$\\
\hline
\multicolumn{2}{c|}{PGSTAT/d.o.f} & 299.2/361.0 & 311.6/361.0 & 312.5/361.0 & 337.2/361.0 & 298.1/361.0 & 303.3/361.0\\
\hline
\hline
\end{tabular}
\label{tabel1}
\end{table*}

\begin{figure*}
\centering
\subfigure{
\begin{minipage}[b]{0.3\textwidth}
\includegraphics[width=1\textwidth]{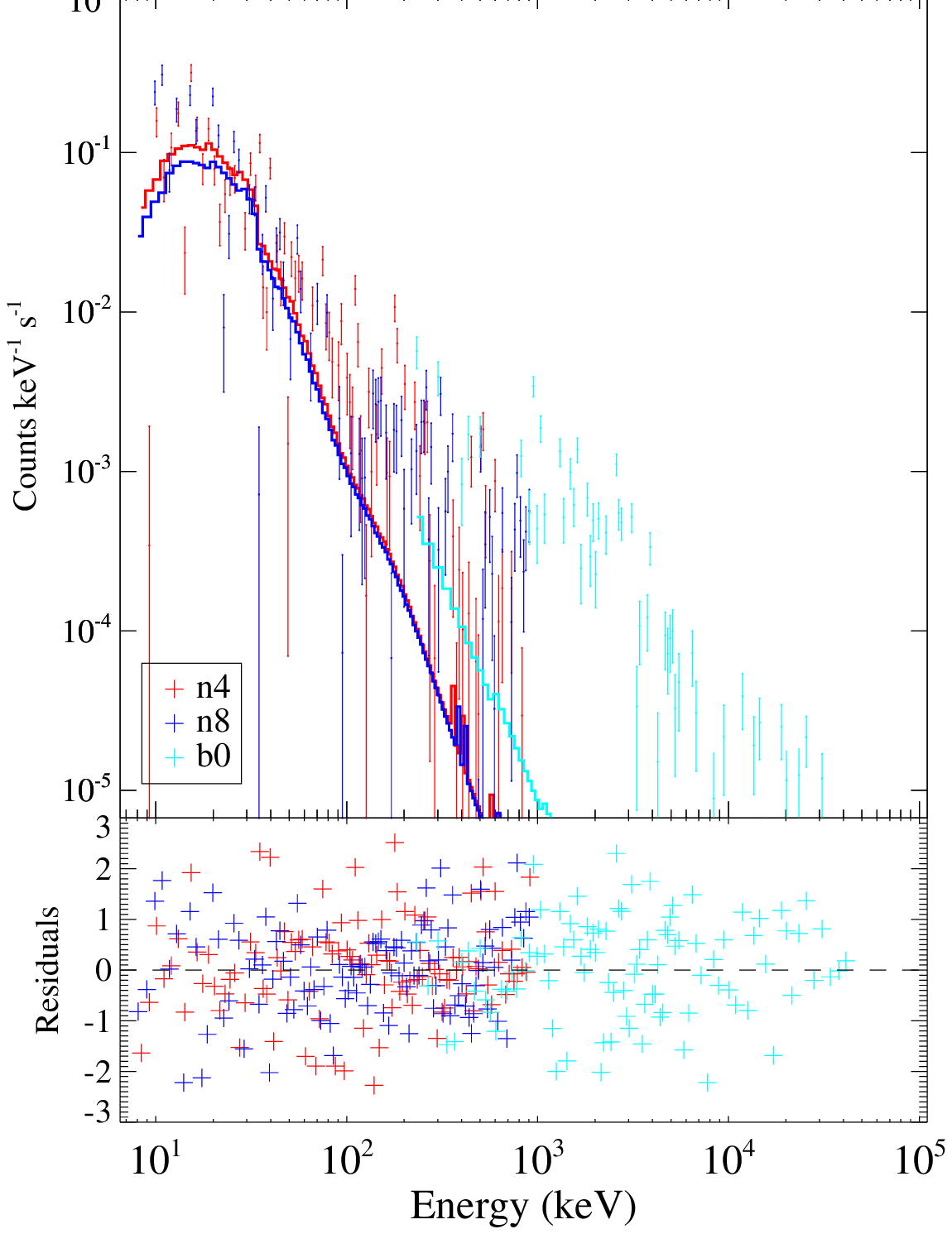}
\label{fig3:subfig:0}
\end{minipage}}
\subfigure{
\begin{minipage}[b]{0.3\textwidth}
\includegraphics[width=1\textwidth]{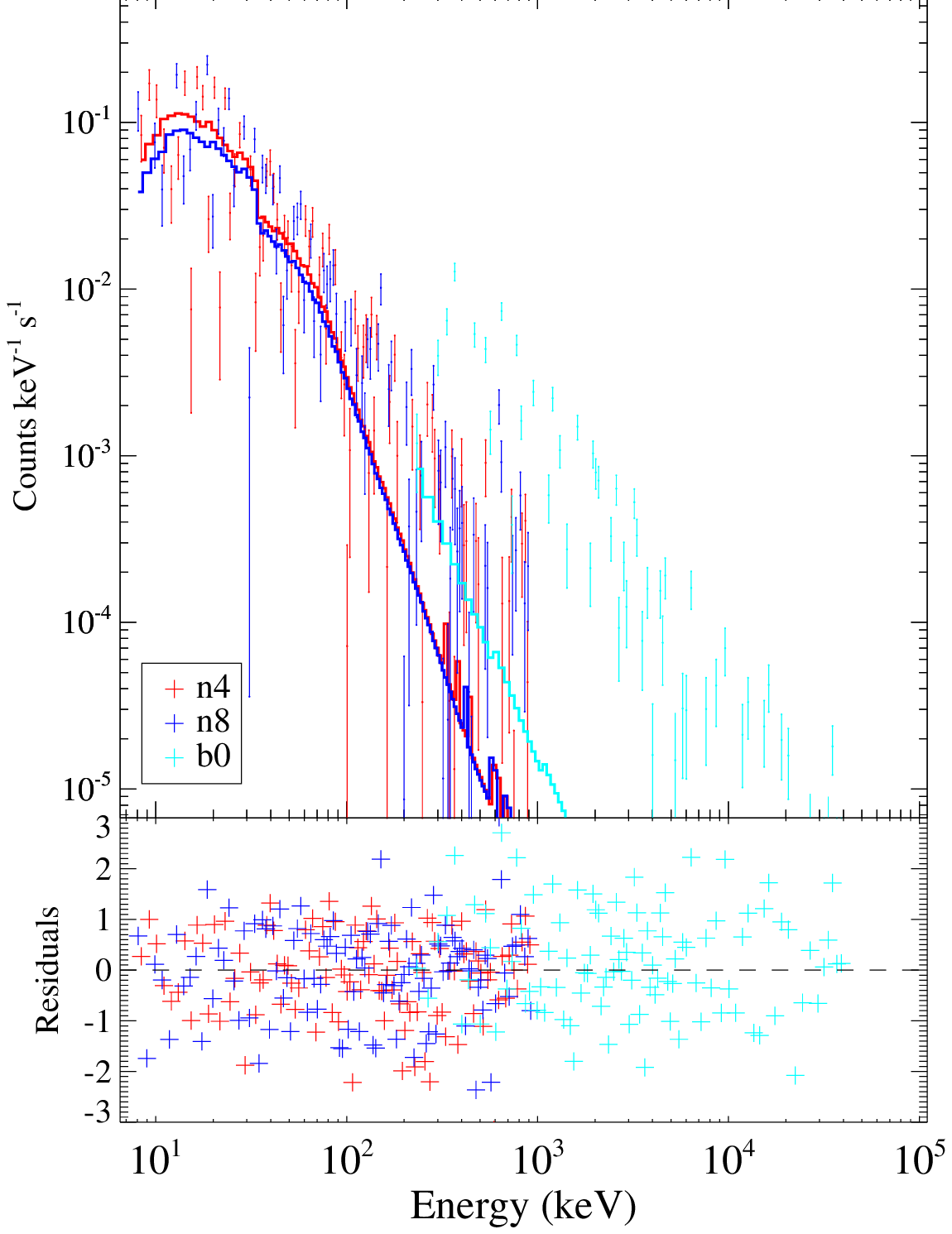}
\label{fig3:subfig:1}
\end{minipage}}
\subfigure{
\begin{minipage}[b]{0.3\textwidth}
\includegraphics[width=1\textwidth]{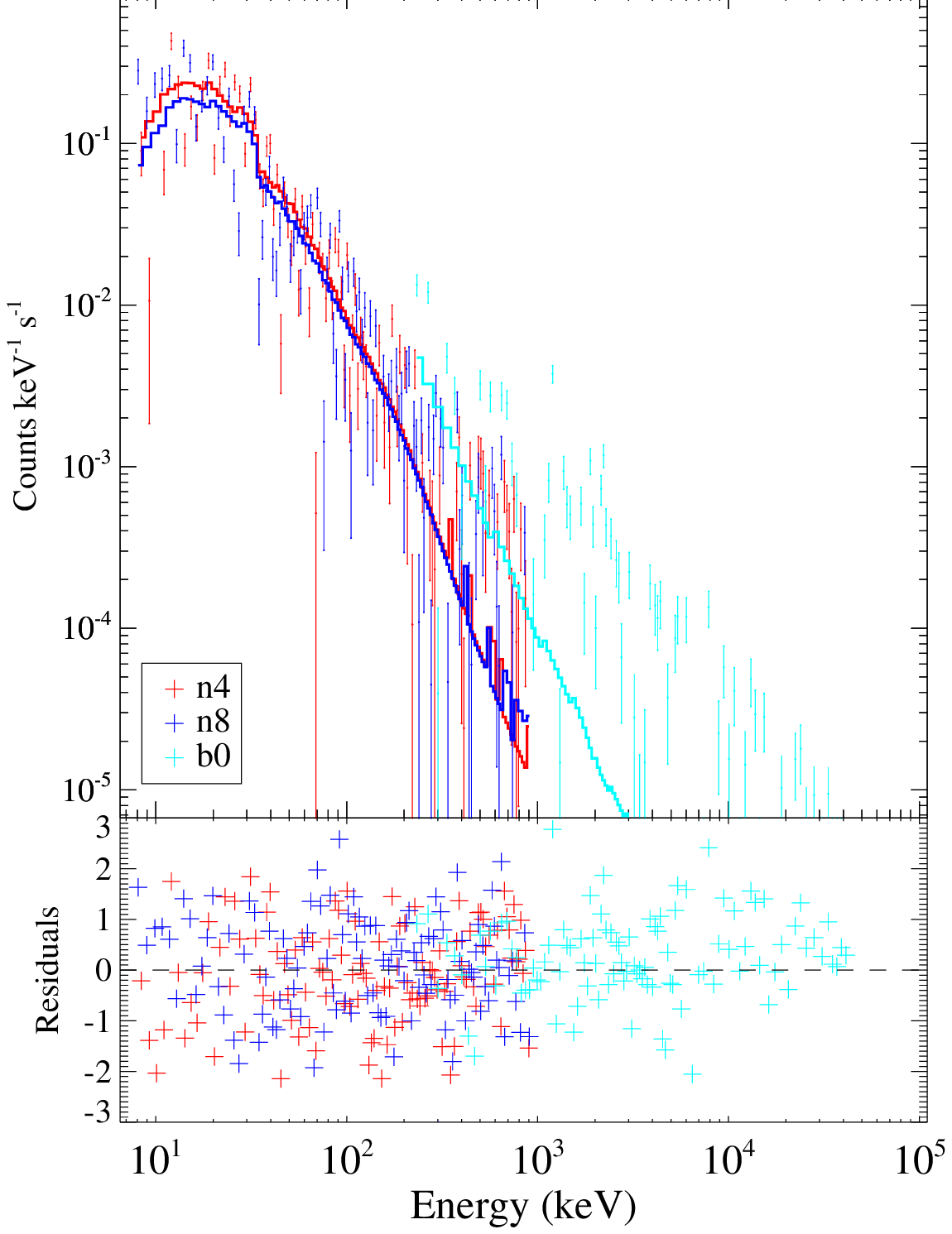}
\label{fig3:subfig:2}
\end{minipage}}
\subfigure{
\begin{minipage}[b]{0.3\textwidth}
\includegraphics[width=1\textwidth]{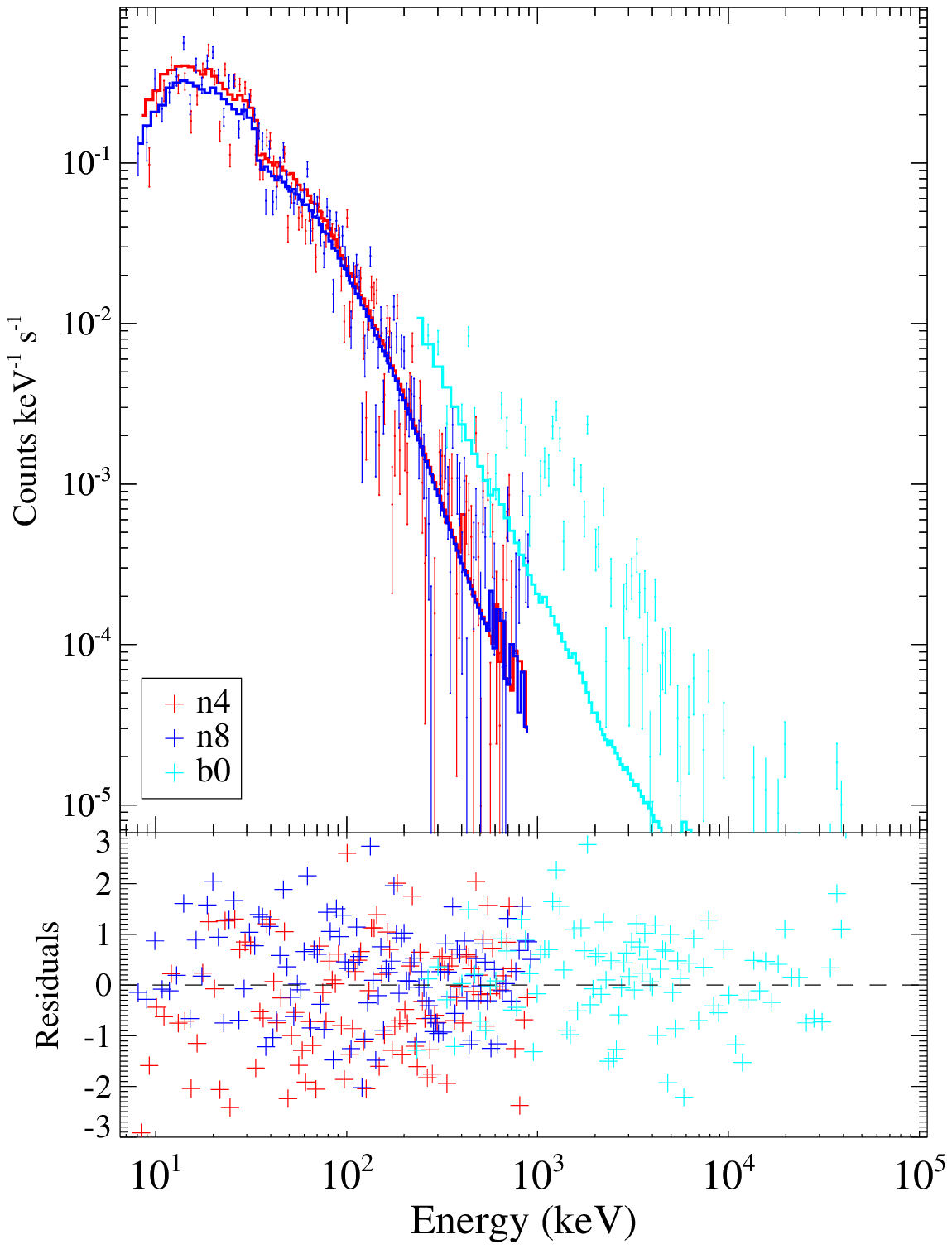}
\label{fig3:subfig:3}
\end{minipage}}
\subfigure{
\begin{minipage}[b]{0.3\textwidth}
\includegraphics[width=1\textwidth]{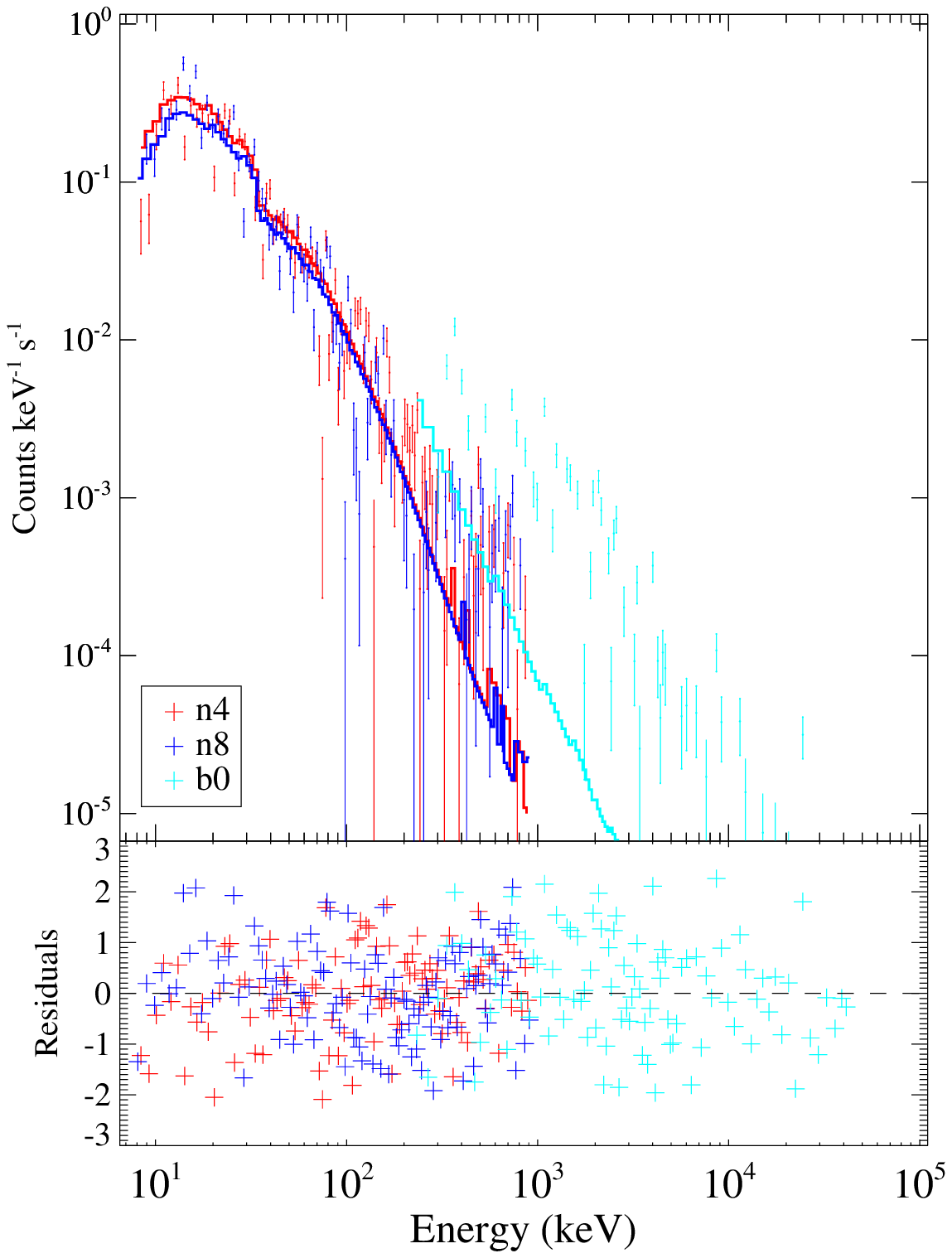}
\label{fig3:subfig:4}
\end{minipage}}
\subfigure{
\begin{minipage}[b]{0.3\textwidth}
\includegraphics[width=1\textwidth]{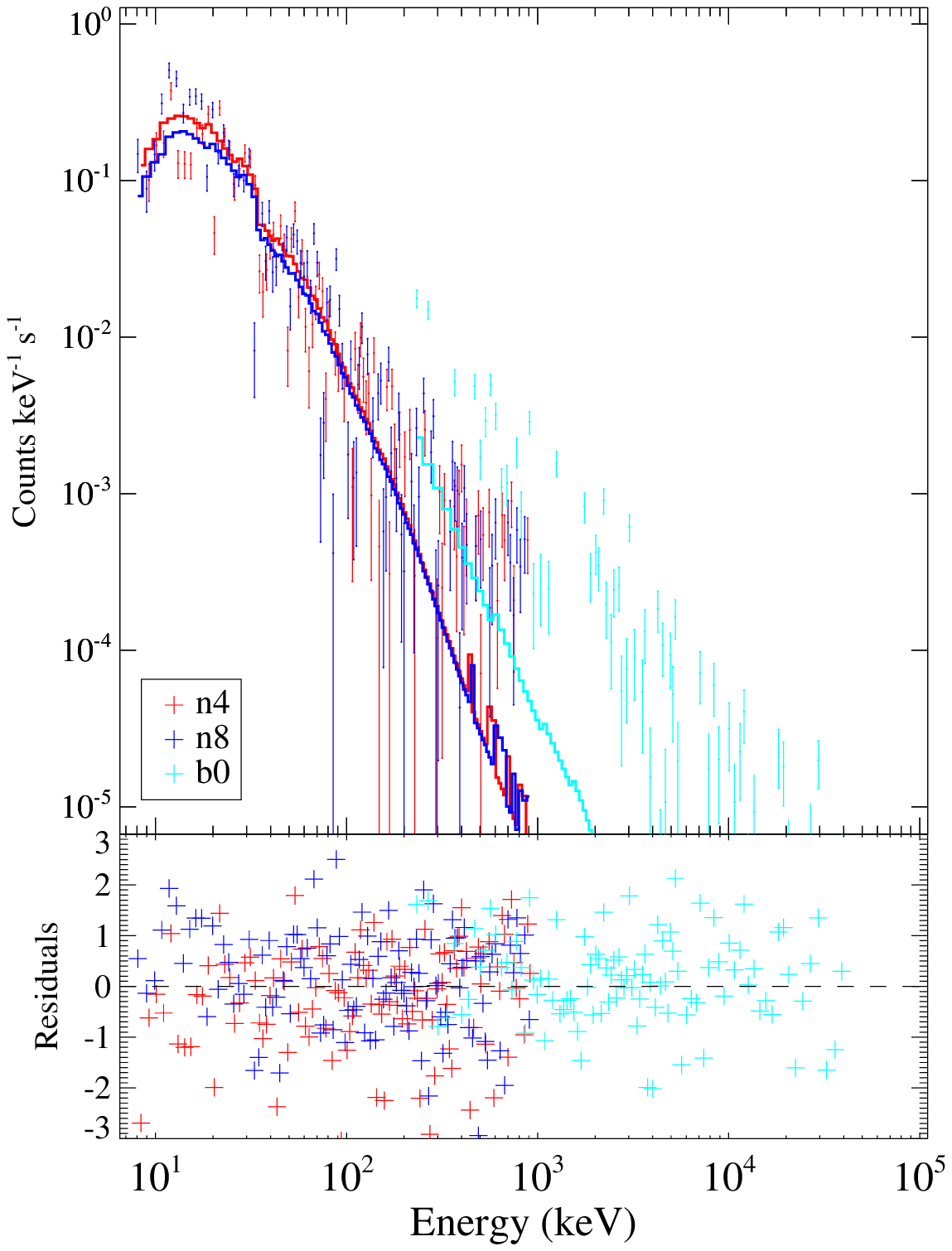}
\label{fig3:subfig:5}
\end{minipage}}
\caption{The observed count spectrum and model count spectrum for six time intervals. n4, n8, b0 represent three {\it Fermi}/GBM detectors. From left to right: first row - intervals 0, 1, 2; second row - 3, 4, 5.\label{fig3}}
\end{figure*}

\begin{figure*}
\centering
\subfigure{
\begin{minipage}[b]{0.3\textwidth}
\includegraphics[width=1\textwidth]{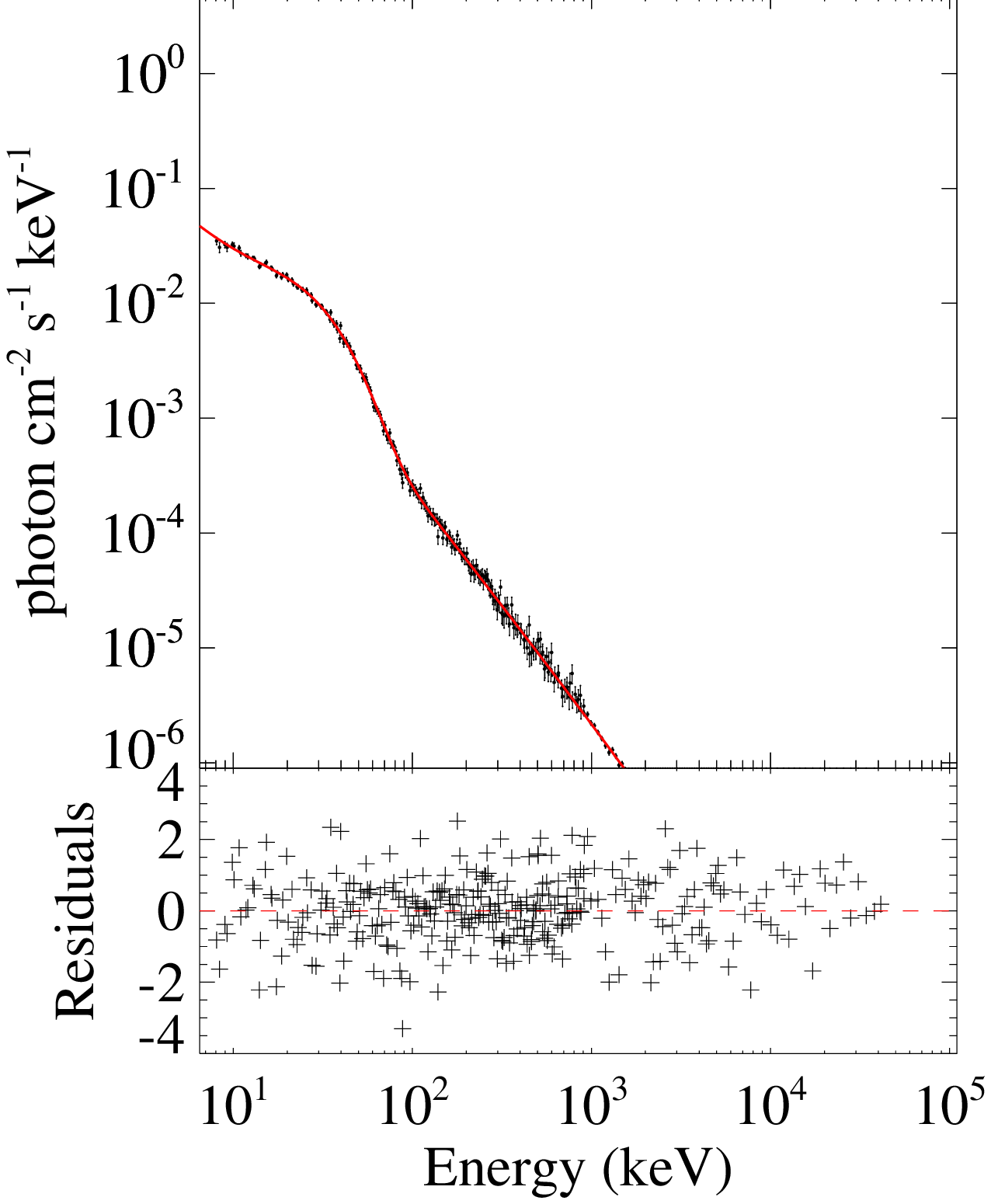}
\label{fig4:subfig:0}
\end{minipage}}
\subfigure{
\begin{minipage}[b]{0.3\textwidth}
\includegraphics[width=1\textwidth]{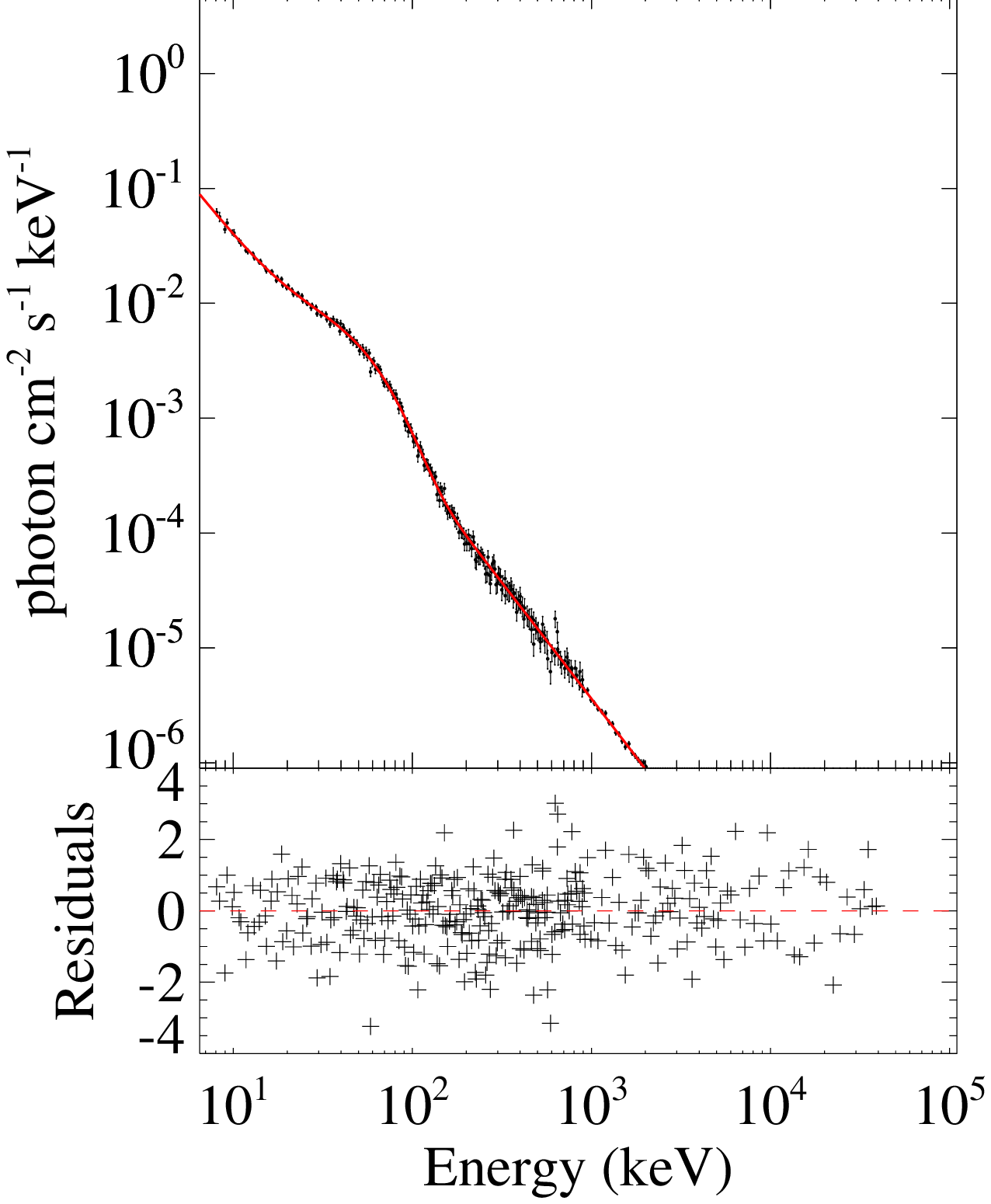}
\label{fig4:subfig:1}
\end{minipage}}
\subfigure{
\begin{minipage}[b]{0.3\textwidth}
\includegraphics[width=1\textwidth]{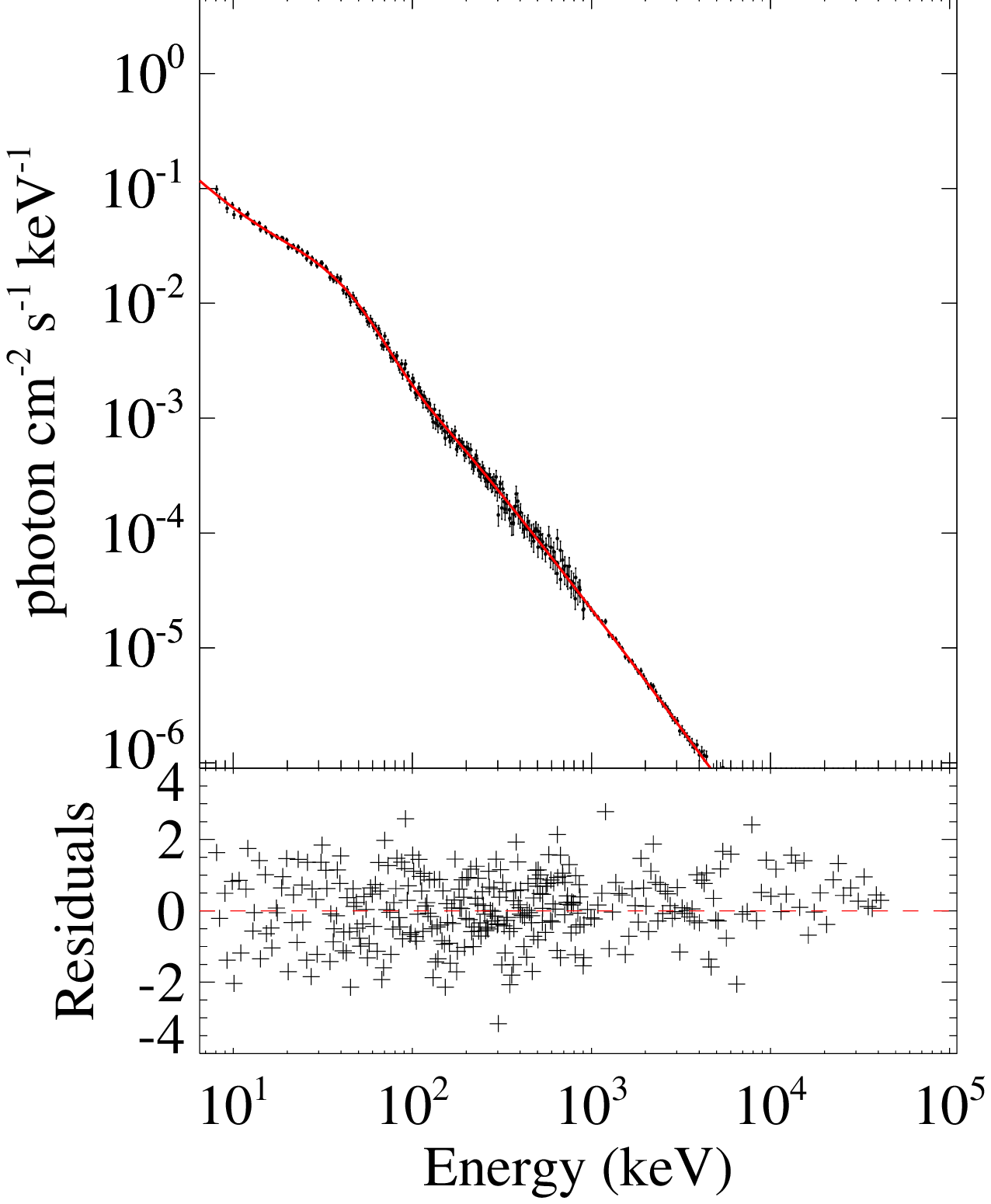}
\label{fig4:subfig:2}
\end{minipage}}
\subfigure{
\begin{minipage}[b]{0.3\textwidth}
\includegraphics[width=1\textwidth]{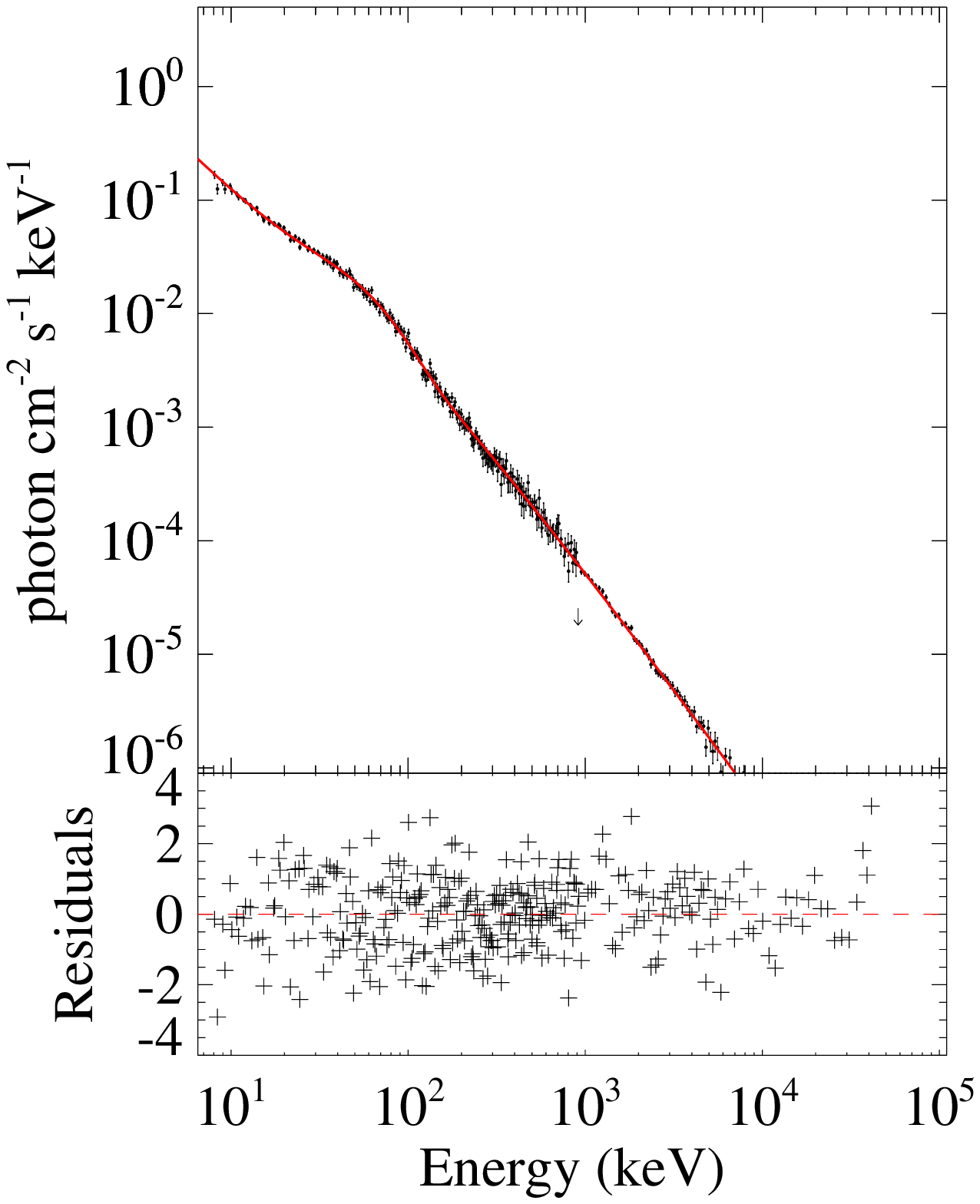}
\label{fig4:subfig:3}
\end{minipage}}
\subfigure{
\begin{minipage}[b]{0.3\textwidth}
\includegraphics[width=1\textwidth]{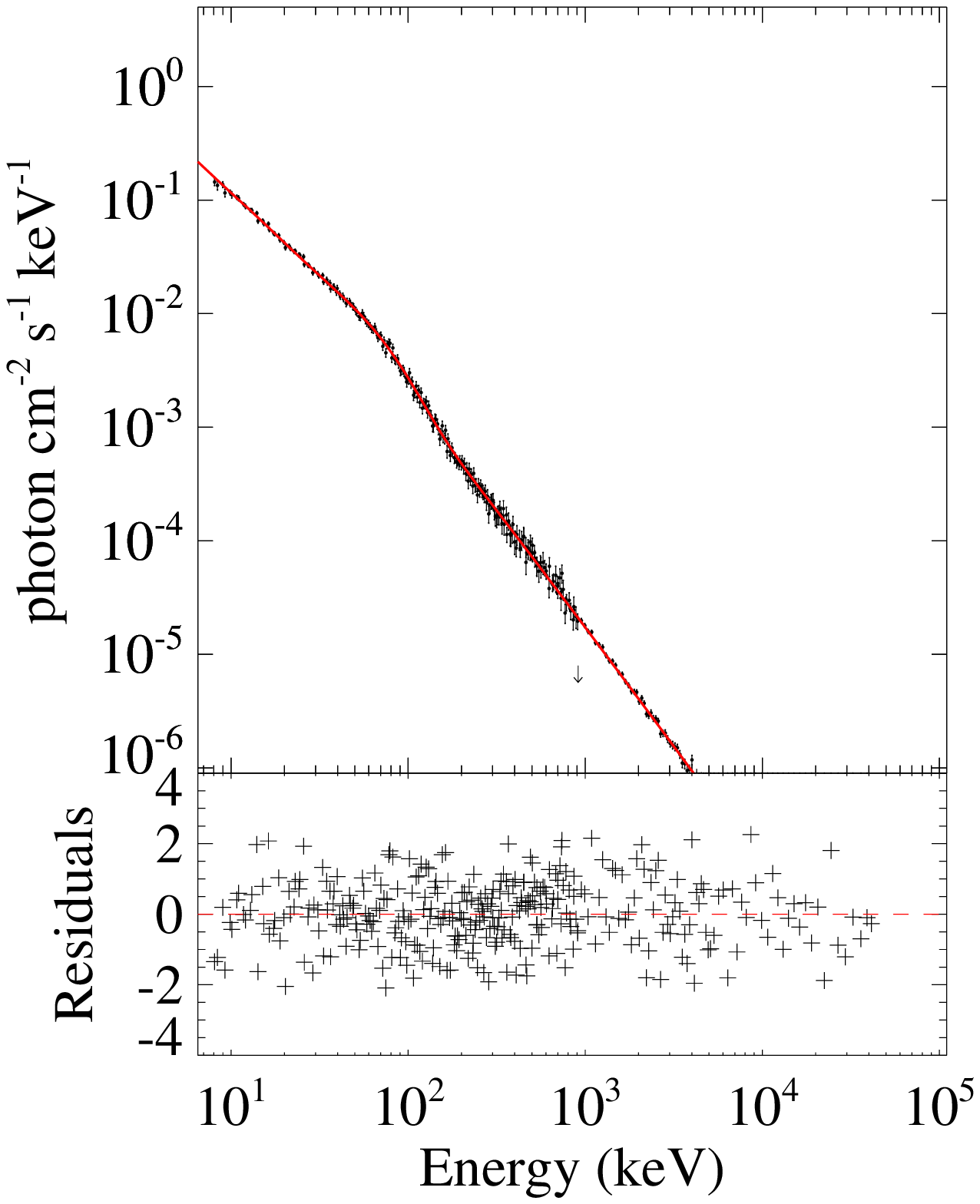}
\label{fig4:subfig:4}
\end{minipage}}
\subfigure{
\begin{minipage}[b]{0.3\textwidth}
\includegraphics[width=1\textwidth]{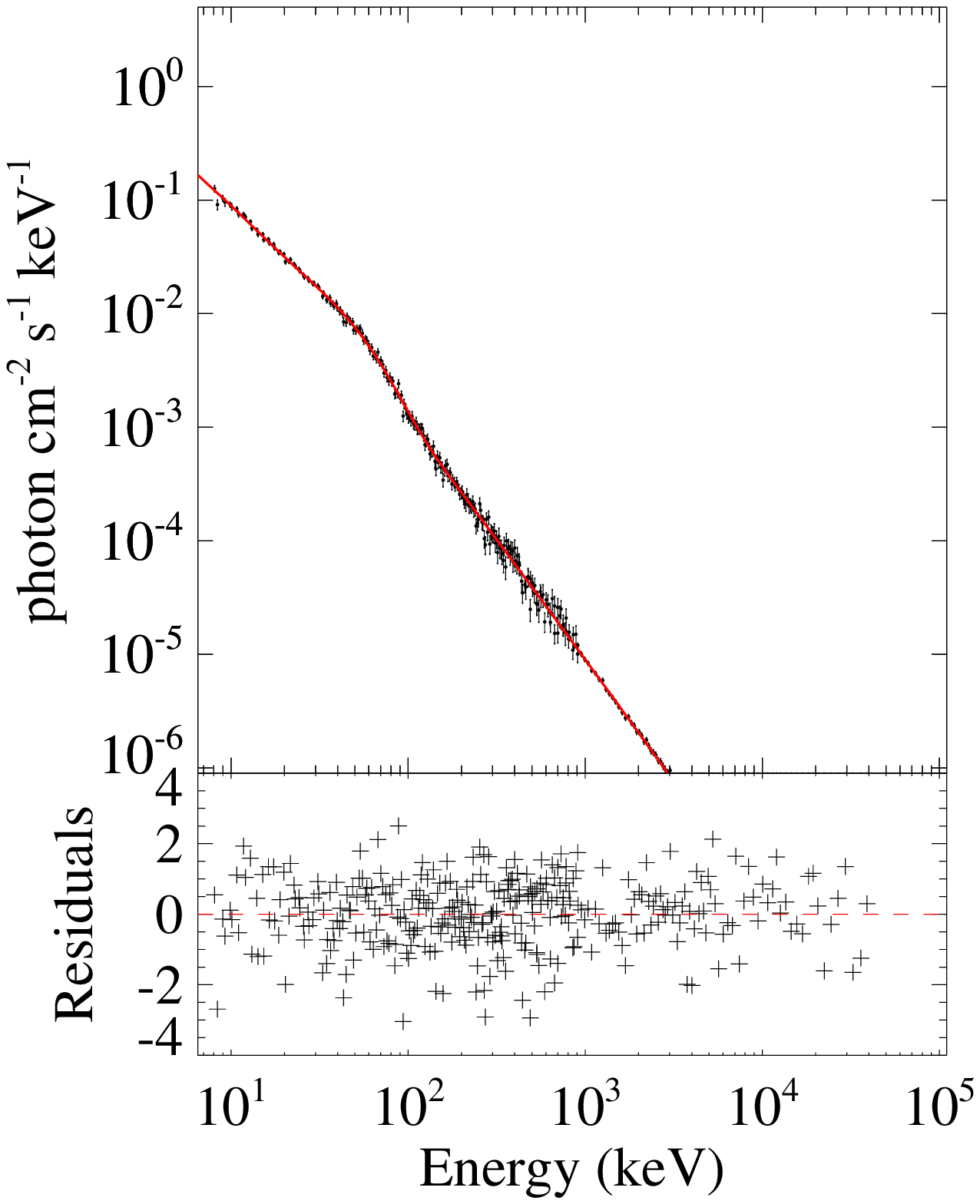}
\label{fig4:subfig:5}
\end{minipage}}

\caption{Same as Figure 3 but for the observed photon flux (black datapoints) and theoretical photon spectrum (red line).
\label{fig4}}
\end{figure*}

\begin{figure}
\label{fig5}
\begin{center}
\includegraphics[width=0.5\textwidth]{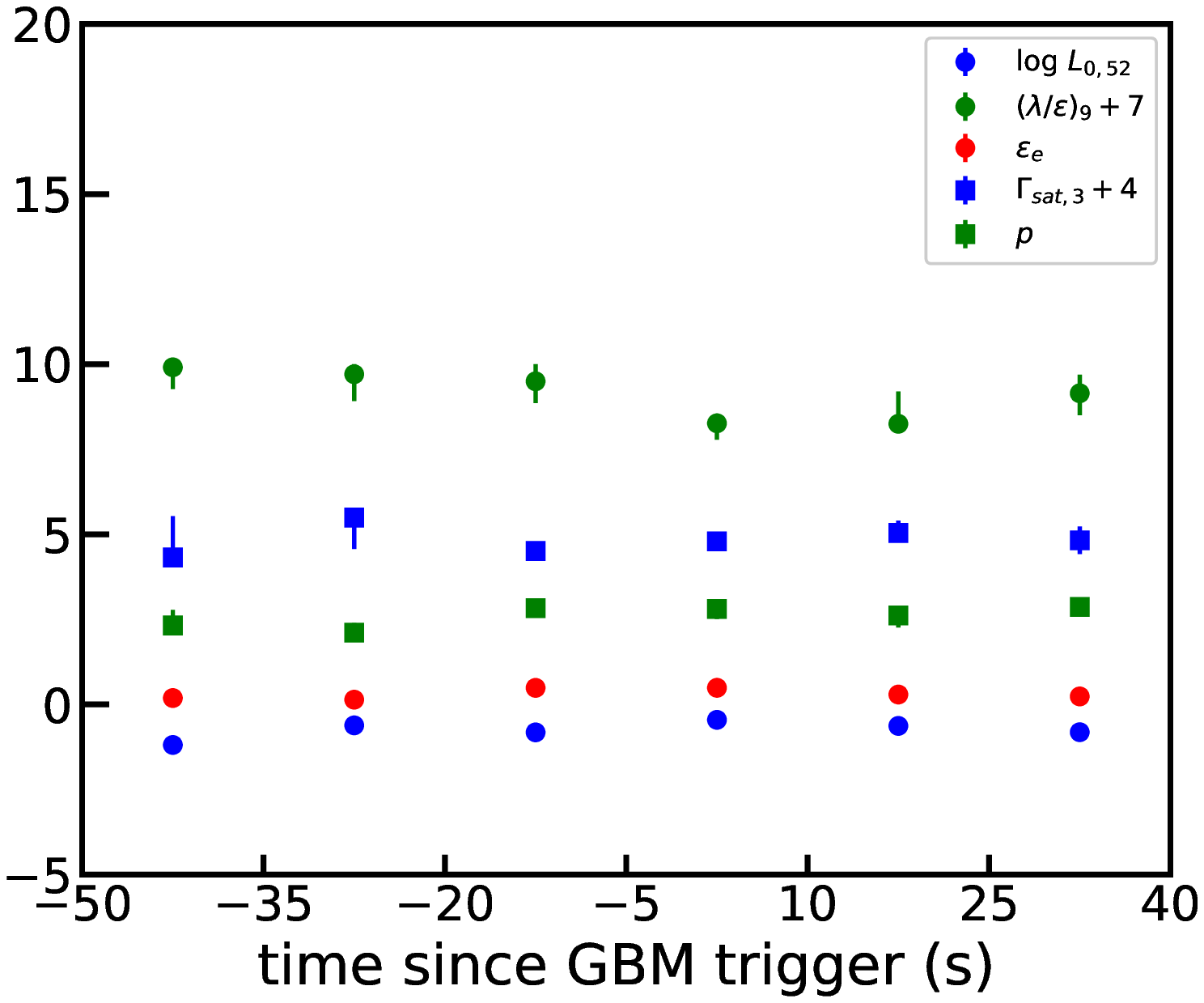}
\caption{The time evolution of best-fitting parameters with the MIGMAD model for GRB 160804A. To avoid possible overlapping, the value of $(\lambda/\epsilon)_9$ is added with 7 and $\Gamma_{\rm sat,3}$ is added with 4 in the figure.}
\end{center}
\end{figure}

\subsection{X-ray Afterglow}
The X-ray afterglow data of GRB 160804A was recorded 137 $\rm s$ after the BAT trigger, and it shows a prominent shallow decay phase without obvious X-ray flares \citep{dav16}. The shallow decay of X-ray afterglow is commonly ascribed to the energy injection from a central magnetar \citep{dai98a,dai98b,zhang01}. We follow the standard procedure described in \citet{yu09,yu10}, in which the X-ray light curve after the initial steep decay phase was fitted by a smoothed double broken power-law function,
\bea
F_X(t)=F_{X,b}\left[\left(\frac{t}{T_{b1}}\right)^{\omega\alpha_1}
+\left(\frac{t}{T_{b1}}\right)^{\omega\alpha_2}\right.\nonumber\\
+\left.\left(\frac{T_{b2}}{T_{b1}}\right)^{\omega\alpha_2}
\left(\frac{t}{T_{b2}}\right)^{\omega\alpha_3}\right]^{-1/\omega},
\ena
where typical $w=3$ is assumed \citep{lia07}. The fitting is performed using McEasyFit \citep{zhang16} and the result is shown in Figure 6, with the best-fitting values being listed in Table 2. In view of the late-time upper limit from {\it Swift}-XRT, a double power-law function generally fits better than a single broken power-law one. This fitting could also have some implications for the magnetar and will be discussed in the following section.

\begin{figure}
\label{fig6}
\begin{center}
\includegraphics[width=0.5\textwidth]{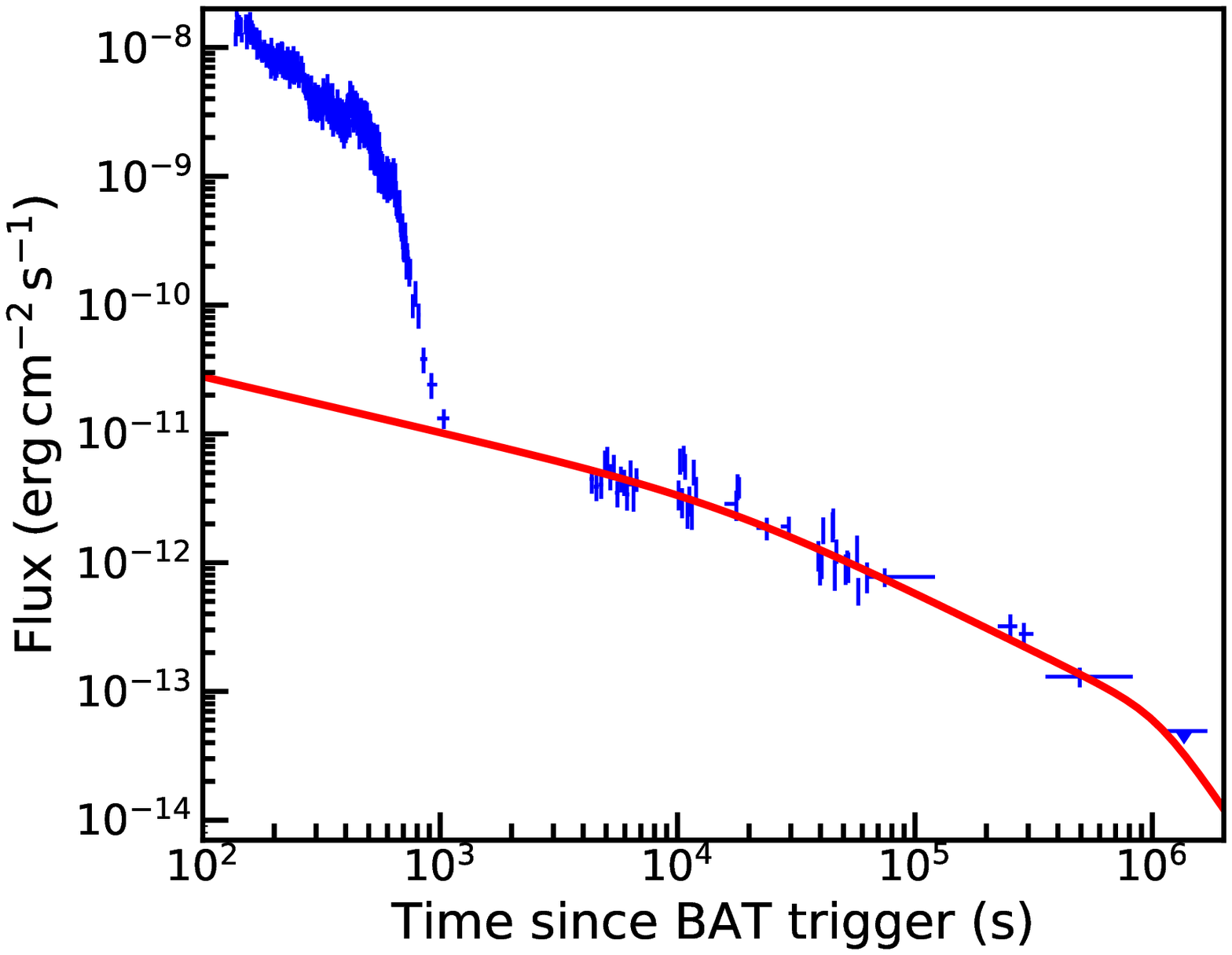}
\caption{Fitting the X-ray afterglow light curve of GRB 160804A (blue datapoints) with double broken power law (red line) function. A second break is needed to reconcile the last upper limit from observation.}
\end{center}
\end{figure}

\begin{table}
\centering
\caption{The best-fitting values for the six parameters defined in equation (12).}
\label{table2}
\begin{tabular}{ccc}
\toprule Parameter &Allowed range & Best-fitting value \\
\midrule$F_{X,b}$ &[$10^{-15},\,10^{-10}$]& $3.09_{-1.69}^{+1.14}\times10^{-12}$ \\
$\alpha_1$ &[-1.0, 1.0] & $0.91_{-0.48}^{+0.02}$ \\
$\alpha_2$ &[0.0, 2.0] & $0.43_{-0.03}^{+0.51}$ \\
$\alpha_3$ &[2.0, 3.0]& $2.71_{-0.55}^{+0.11}$ \\
$T_{b1}$ &[$10^3,\,10^5$]& $1.62_{-0.63}^{+3.03}\times10^4$ \\
$T_{b2}$ &[$10^4,\,10^6$]& $4.40_{-0.70}^{+4.32}\times10^5$\\
\bottomrule & &
\end{tabular}%
\end{table}

\subsection{Implications for the Central Magnetar}
Firstly, for a double power-law type afterglow, the two break times are believed to be correlated to the magnetar spin evolution \citep{yu10}. Before the first break, the gravitational wave radiation dominates the spin-down and its typical timescale is $T_g\sim T_{b1}/(1+z)$. For GRB 160804A, we get $T_g\simeq9.35_{-3.62}^{+19.05}\times10^3\,\rm s$. The other break time is attributed to a break of the magnetic dipole luminosity at $T_c\sim T_{b2}/(1+z)\simeq2.53_{-0.40}^{+2.49}\times10^5\,\rm s$ and the typical magnetic braking timescale is $T_m=(T_gT_c)^{1/2}\simeq4.87_{-1.37}^{+7.08}\times10^4\,\rm s$. Since the initial period and magnetic field is given by
\bea
P_{i,-3}=a^{-1/6}T_{g,3}^{1/6},\nonumber\\
B_{i,14}=1.4P_{i,-3}T_{m,5}^{-1/2},
\ena
where $a$ is a slow-varying numerical factor of order unity \citep{yu09, yu10}, we can deduce that $P_i=1.45_{-0.11}^{+2.30}\,\rm ms$ and $B_i=2.91_{-0.68}^{+0.26}\times10^{14}\,\rm G$ for the magnetar of GRB 160804A.

Secondly, we can get some information of the central magnetar from the prompt emission. In the prompt phase ($T_{90}\ll T_g$), gravitational braking dominates the spin-down. Since the energy loss rate of gravitational radiation is in proportion to the six power of angular velocity \citep{sha83}, the evolution of angular velocity can be obtained by solving the spin evolution equation, which gives
\beq
\Omega(t)=\Omega_i\left(1+\frac{4\Omega_i^4}{\Omega_s^4}\frac{(t+t_0)}{T_g}\right)^{-1/4},
\enq
where $\Omega_i, \Omega_s$ represent the initial, the stable angular velocity after gravitational braking respectively. We introduce a parameter $t_0$ here to account for the possible time delay between the birth of magnetar and the GRB trigger. This is naturally expected since the magnetar should cool first and jet break-out also takes some time \citep{met11}. The spin period is then $P(t)=2\pi/\Omega(t)$ and the evolution of parameter $(\lambda/\epsilon)$ can be obtained as constant $\epsilon=0.1$ is assumed. Moreover, the magnetic dipole luminosity is $L_m(t)=B^2R^6\Omega^4/6c^3$. If the magnetic field and jet beaming factor $f_B=(1-\cos\theta_j)\simeq\theta_j^2/2$ do not vary with time, theoretically we have $L_0(t)=(1/4\pi)L_m/f_B=(1/4\pi)B^2R^6/(6c^3f_B)\times\Omega(t)^4\equiv K\Omega(t)^4$. Taking $\Omega_i,\,\Omega_s,\,t_0,\,K$ as parameters, we can perform a combined fit of time evolution for parameters $L_0$ and $(\lambda/\epsilon)$ that obtained in Section 3.1. The fitting results are shown in Figure 7 and Table 3. As we can see, the initial spin period deduced from prompt emission fitting is $P_i=2\pi/\Omega_i=1.43_{-0.30}^{+0.27}\,\rm ms$, which matches perfectly with the value of $P_i=1.45_{-0.11}^{+2.30}\,\rm ms$ from afterglow fitting in section 3.2. These two independent approaches giving a consistent result adds great credibility to the MIGMAD model and a millisecond magnetar central engine. Note that the theoretical curves in Figure 7 are obtained just through the time evolution of spin period, while \citet{ben17b} considered time evolutions of several other quantities (e.g, dipole field strength, baryon loading, etc).

Lastly, more details can be deduced if we further do a comprehensive analysis. On one hand, equation (13) gives a best-fitting magnetic field $B_i=2.91\times10^{14}\,\rm G$. On the other hand, from equation (1) we can get that magnetic field $B\propto r^{-4/3}$ in the Poynting-flux dominated jet. Taking time interval-3 (main peak) as a representative, tracing back to a position near the surface of the magnetar, we can get $B(r)=6.13\times10^{14}r_6^{-4/3}\,\rm G$\footnote{Strictly speaking, the Poynting-flux dominated outflow originate near the light cylinder \citep{aha12}, at a radius $R_L=c/\Omega\simeq4.78\times10^6\,\rm cm$ for a typical millisecond magnetar. However, the dipole magnetic field decays with a radius follows $B\propto r^{-b}$ where b is between 1 (toroidal) and 2 (poloidal). Thus, tracing back to light cylinder or the magnetar surface will not make a significant difference in the calculation.}. Comparing with the field strength from afterglow fitting, the radius of the magnetar can be determined as $R=1.75\times10^6\,\rm cm$. Since $K\equiv B^2R^6/(6c^3f_B)$, substituting the best-fitting value of $K$ and $R$ we can get $f_B=2\times10^{-6}$, which corresponds to a narrowly beamed jet with half opening angle $\theta_j\simeq0.002$. This value is much smaller than the opening angles that are usually deduced from afterglow jet-break signatures. On one hand, for these extremely narrow jets, the jet-break may happen at very early times \citep[e.g.][]{frail01}, which are not easy to be identified since the X-ray afterglows may be still in a steep decay phase. This may explain why so many afterglows do not show jet-break signatures. In this sense, observationally there may exist a bias, since jet-break phenomena are easier to be identified for jets with larger opening angles. On the other hand, for a magnetized jet, during propagating in the progenitor envelope, the jet head will contract due to strong hoop stress, which gives rise to a nozzle-like jet shape after breakout \citep{bro14}. Basically, the opening angle of the magnetized jet could be much smaller than that of the hydrodynamic jet. In the recent numerical simulation work \citep{bro16} we can see that the magnetized jet could be very ¡°slim¡± (Figure 15 in their paper). Also, the opening angle depends on the density profile of medium. A denser envelope generally leads to a smaller jet opening angle. In view of these arguments above, the peculiar $\theta_j=0.002$ for this GRB 160804A is still reasonable, though not very typical.

\begin{figure}
\label{fig7}
\begin{center}
\includegraphics[width=0.5\textwidth]{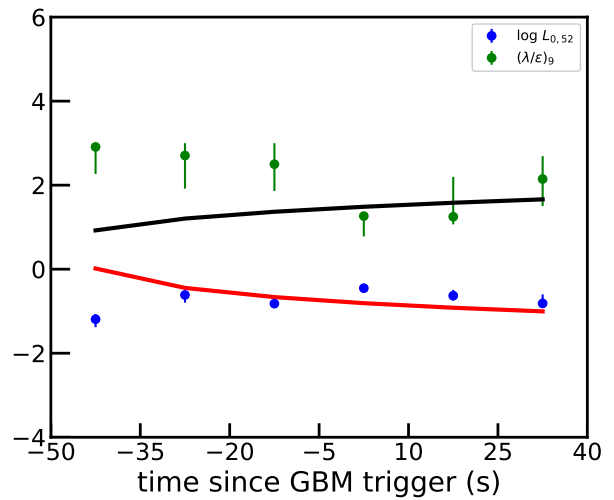}
\caption{The theoretical time evolutions (red line and black line) for the two spectral fitting parameters $L_{0,52}$ (blue points) and $(\lambda/\epsilon)_9$ (green points) in the MIGMAD model.}
\end{center}
\end{figure}

\begin{table}
\centering
\caption{The best-fitting values for the four parameters defined in equation (14).}
\label{table3}\center%
\begin{tabular}{ccc}
\toprule Parameter &Allowed range & Best-fitting value \\
\midrule$\Omega_i$ &[$3\times10^3,\,6\times10^3$]& $4.40_{-0.69}^{+1.18}\times10^3$ \\
$\Omega_0$ &[6, 600] & $492.16_{-7.78}^{+14.22}$ \\
$t_0$ &[50, 150] & $50.04_{-0.03}^{+1.54}$ \\
$K$ &[$10^{35},\,6\times10^{38}$]& $5.98_{-0.13}^{+0.01}\times10^{38}$ \\
\bottomrule & &
\end{tabular}%
\end{table}

\section{Discussions and Conclusions}
In this work we have revisited the scenario for the prompt emission of GRB, called the MIGMAD model, by assuming that the central engine is a newborn millisecond magnetar, whose wind undergoes an internal gradual magnetic dissipation. This model is relatively clean and has only few assumptions. Theoretically, we have investigated the high-energy emission from the wind of this newborn magnetar, which should exist as long as the wind is initially Poynting-flux dominated. We have applied it to the prompt emission of GRB 160804A and showed that its spectrum can be fitted very well. We found that a beamed wind with half opening angle 0.01 from a newborn magnetar that has an initial spin period 1.45 ms and surface magnetic field $2.91\times10^{14}\,\rm G$ can reproduce the prompt emission and afterglow behaviors well. In addition, we can also predict a subclass of GRBs that are produced based on this scenario, which are named as ``MIGMAD bursts".

MIGMAD bursts should have distinctive properties compared to normal ones. First, a general testable prediction of the MIGMAD model is that the peak of the flux spectrum determined by the photospheric component is usually around a hundred keV and not so sensitive to the model parameters \citep{ben17}. Second, the most prominent property is that the durations of MIGMAD bursts might be very long (up to $\sim10^2-10^4\,\rm s$), depending mainly on the spin-down timescale of the central magnetar. Actually, we know little about the production of ultra-long GRBs (ULGRBs) now. Existing models for ULGRBs include collapses of blue supergiant stars \citep{gen13,lev14}, newborn magnetars \citep{gre15, met18}, white dwarf tidal disruption events \citep{iok16} and black hole hyper-accretion processes \citep{liu18}. Especially, \citet{gre15} claimed that the associated supernova (SN 2011kl) of ULGRB 111209A is powered by a magnetar. However, the radiation mechanism of ULGRB has not been detailedly discussed in their work. In this work we show that MIGMAD scenario is able to explain the prompt emission properties, making the magnetar origin model concrete and complete. In this sense, we might already have some other interesting candidates like GRB 111209A and GRB 060218, which will be investigated in our future work. Once the deduced spin period and magnetic field can match well from the independent analysis of prompt emission, afterglow and associated supernova observations for one GRB, its magnetic central engine would be finally fully convinced.

Due to the limited available spin-down luminosity of a millisecond magnetar and the low efficiency in converting magnetic energy to radiation by gradual dissipation \citep{xiao17b}, given the current {\it Fermi}-GBM sensitivity we can only detect relatively nearby MIGMAD bursts with redshift less than 1. Even though, {\it Fermi}-GBM should have caught a few events so that it is meaningful to look for candidates in the archival data. More facilities with better sensitivities (e.g. {\it Insight}-HXMT) would be able to detect other more MIGMAD bursts in the near future.

\acknowledgements
This work was supported by the National Basic Research Program of China (973 Program, Grant No. 2014CB845800), the National Key Research and Development Program of China (Grant No. 2017YFA0402600) and the National Natural Science Foundation of China (Grant No. 11573014). DX was also supported by the Natural Science Foundation for the Youth of Jiangsu Province (Grant NO. BK20180324) and the Fundamental Research Funds for the Central Universities (Grant No. 020114380027). BBZ acknowledge the support from the National Thousand Young Talents program of China. We acknowledge the use of the public data from the {\it Fermi} satellite.

%\clearpage

\clearpage

\end{document}